\documentstyle[prd,aps,epsf,graphicx]{revtex}

\DeclareFontFamily{OT1}{rsfs10}{}
\DeclareFontShape{OT1}{rsfs10}{m}{n}{ <-> rsfs10 }{}
\DeclareMathAlphabet{\mathscript}{OT1}{rsfs10}{m}{n}

\newcommand{\be}{\begin{equation}}
\newcommand{\ee}{\end{equation}}
\newcommand{\nn}{\nonumber}
\newcommand{\bea}{\begin{eqnarray}}
\newcommand{\eea}{\end{eqnarray}}
\newcommand{\ba}{\begin{array}}
\newcommand{\ea}{\end{array}}

\newcommand{\ns}{\normalsize}
\newcommand{\pt}{\partial}

\newcommand{\eqref}[1]{(\ref{#1})}

\def\a{\alpha}
\def\b{\beta}

\def\c{\chi}
\def\d{\delta}
\def\e{\epsilon}

\def\f{\phi}
\def\z{\psi}

\def\k{\kappa}

\def\m{\mu}
\def\n{\nu}

\def\p{\pi}

\def\r{\rho}
\def\s{\sigma}

\def\t{\tau}

\def\z{\zeta}

\def\F{\Phi}

\def\bx{{\bf x}}
\def\bal{{\mbox{\boldmath $\alpha$}}}
\def\bq{{\bf q}}

\def\bd{{\bf d}}

\def\bp{{\bf p}}
\def\bH{{\bf H}}

\begin{document}

\begin{titlepage}

\title{
\hfill{\ns SUSX-TH/01-029\\}
\hfill{\ns hep-th/0106285\\[2cm]}
{\LARGE Moving Five-Branes in Low-Energy Heterotic M-Theory}\\[1cm]}
\setcounter{footnote}{0}
\author{{\ns\large
 Edmund J.~Copeland\footnote{email: e.j.copeland@sussex.ac.uk}~,
\setcounter{footnote}{3}
 James Gray\footnote{email: kapm3@pact.cpes.sussex.ac.uk} and
 Andr\'e Lukas\footnote{email: a.lukas@sussex.ac.uk}} \\[0.8em]
      {\ns Centre for Theoretical Physics,
     University of Sussex}\\[-0.2em]
      {\ns Falmer, Brighton BN1 9QJ, United Kingdom}}


\maketitle

\vspace{1cm}

\begin{abstract}
We construct cosmological solutions of four-dimensional effective
heterotic M-theory with a moving five-brane and evolving dilaton
and $T$ modulus. It is shown that the five-brane generates
a transition between two asymptotic rolling-radii solutions.
Moreover, the five-brane motion always drives the solutions towards
strong coupling asymptotically. We present an explicit example
of a negative-time branch solution which ends in a brane collision
accompanied by a small-instanton transition. The five-dimensional
origin of some of our solutions is also discussed.
\end{abstract}

\thispagestyle{empty}

\end{titlepage}


\section{Introduction}

Brane-world models share a number of characteristic properties which
make them a particularly interesting starting point for early universe
cosmology. One common feature which has stimulated most of the work on
brane-world 
cosmology~\cite{Lukas:1999qs,Randall:1999vf,Lukas:2000yn,Binetruy:2000hy,Csaki:2000mp,Ida:2000ui,Hawking:2000kj,Carter:2001nj,Deffayet:2001uy} to 
date is the presence of stress energy
localised on the brane leading to inhomogeneous additional
dimensions. In this paper, we will be concerned with another generic
property, namely the possibility that branes move as part of the
cosmological evolution. While there have been some early proposals
for moving brane cosmological scenarios~\cite{Dvali:1999pa} the problem has
only recently received wider 
attention~\cite{Khoury:2001wf,Kallosh:2001ai,Burgess:2001fx,Khoury:2001iy,Kallosh:2001du}.

\vspace{0.4cm}

Here, we will not be so much concerned with cosmological scenarios but
rather with constructing and analysing fundamental cosmological
solutions with moving branes in a well-defined M-theory context. More
specifically, we will consider heterotic
M-theory~\cite{Witten:1996mz,Horava:1996qa,Horava:1996ma} compactified
on a Calabi-Yau three-fold. The resulting five-dimensional theory
consists of an $N=1$ bulk supergravity theory on the orbifold
$S^1/Z_2$ coupled to two four-dimensional $N=1$ theories residing on
branes which are stuck to the two orbifold
planes~\cite{Lukas:1999yy,Lukas:1999tt,Lukas:2000nh}. In addition,
M-theory five-branes can be included in the vacuum of the
11-dimensional
theory~\cite{Witten:1996mz,Lukas:1999hk,Derendinger:2001gy}. These
five-branes are transverse to the orbifold direction and wrap complex
cycles in the Calabi-Yau three-fold. Therefore, they appear as
additional three-branes in the effective $D=5$ theory. Unlike the
three-branes on the orbifold fix planes, these additional branes can
move in the orbifold direction and it is this new possibility that we
wish to analyse in a cosmological context. A five-brane may then
collide with one of the boundaries in the course of the cosmological
evolution. It is then converted into an instanton located on this
boundary through a small-instanton
transition~\cite{Witten:1996gx,Ganor:1996mu,Ovrut:2000qi}.

\vspace{0.4cm}

Practically, we will be working in the context of the
four-dimensional effective theory, obtained by further reduction on
the domain-wall vacuum of the $D=5$ theory~\cite{Lukas:1999yy}. The simplest
version of this $D=4$, $N=1$ theory contains three chiral superfields,
namely the dilaton $S$, the universal $T$ modulus and the modulus $Z$
which specifies the position of the five-brane in the orbifold
direction. This effective action constitutes a generalisation of the
standard low-energy action for $E_8\times E_8$ heterotic string to
include the five-bane modulus $Z$. In this paper, we will not explicitly
include a non-perturbative superpotential for these moduli but rather
focus on the simplest possibility of freely rolling fields.

As we will see, the axions in these superfields can be set to zero
consistently. Our problem is then to study the cosmological evolution
of the dilaton $\f$ measuring the Calabi-Yau volume, the real part
of the $T$ modulus $\b$ measuring the orbifold size, the five-brane
position $z$ and the scale factor $\a$ of the universe. It is
immediately clear from the corresponding effective action, that
the fields $\f$ and $\b$ cannot be set to constants as soon as
the five-brane moves. Therefore, the evolution of $z$ cannot be
studied in isolation and we must include the standard moduli
$\f$ and $\b$ in our analysis as well.

\vspace{0.4cm}

Let us briefly summarise our main results. We present a general method
of analysing systems with moving branes which is based on a
cosmological moduli space Lagrangian~\cite{Lukas:1997iq}. This allows
us to find the most general solution for our problem and makes it
particularly simple to incorporate a number of generalisations. As
usual, the solutions we find each exist for the negative as well as
the positive time branch.  Asymptotically, at either end of the
negative or positive semi-infinite time ranges, our solutions reduce
to the standard rolling radii solutions involving $\a$, $\b$ and $\f$
only. The five-brane is practically at rest in these limits. However,
the early and late asymptotic regions correspond to two different
rolling radii solutions, in general.  The transition between these two
solutions is generated by the five-brane which only moves during an
intermediate period. Interestingly, not all rolling-radii solutions
can be obtained asymptotically but only those which lead to strong
coupling. Hence, while there are standard rolling radii solutions
which approach weak coupling at one end of the time range, five-brane
motion precludes this possibility.  Another interesting feature of the
solutions is that the five-brane only moves a finite (coordinate)
distance. For each solution, one can therefore adjust whether or not
the five-brane collides with one of the boundaries. We present an
explicit example, where a negative time-branch solution ends with a
brane collision accompanied by a small-instanton transition. Our
solutions and their generalisations to include the effect of
a potential provide a generalised framework for negative time branch
cosmologies. As such they form the common ground between the
``traditional'' dilaton-driven pre-big-bang
scenario~\cite{Gasperini:1993em} and the related  ``ekpyrotic''
universe scenario~\cite{Khoury:2001wf} where the dilaton
is replaced by the five-brane modulus.

\section{Four-dimensional effective action}

To set the scene, we would first like to present the four-dimensional
low-energy effective theory we will be using and explain its
relationship to the underlying heterotic M-theory. Of particular
importance for the interpretation of our results is the relation to
heterotic M-theory in five dimensions, obtained from the
11-dimensional theory by compactification on a Calabi-Yau
three-fold. This five-dimensional theory provides an explicit
realisation of a brane-world theory.

\vspace{0.4cm}

Our starting point is 11-dimensional Horava-Witten
theory~\cite{Horava:1996qa,Horava:1996ma}, that is 11-dimensional
supergravity on the orbifold $S^1/Z_2\times M_{10}$ coupled to two
10-dimensional $E_8$ super-Yang-Mills theories each residing on one of
the two orbifold fixed planes. Upon compactification of this theory on
a Calabi-Yau three-fold to
five-dimensions~\cite{Lukas:1999yy,Lukas:1999tt} one obtains a
five-dimensional $N=1$ gauged supergravity theory on the orbifold
$S^1/Z_2\times M_4$. This bulk theory is coupled to two $D=4$, $N=1$
theories which originate from the ten-dimensional $E_8$ gauge theories
and are localised on three-branes which coincide with the now
four-dimensional orbifold fixed planes.

An important fact for the present paper is that in compactifying from
11 to five dimensions, additional five-branes can be included in the
vacuum~\cite{Lukas:1999hk,Derendinger:2001gy}. These five-branes are transverse to the
orbifold and wrap (holomorphic) two-cycles within the Calabi-Yau
three-fold.  Hence, they also appear as three-branes in the 
five-dimensional
effective theory. Unlike the ``boundary'' three-branes which are stuck
to the orbifold fix points, however, these three-branes are free to move
in the orbifold direction. Altogether, the five-dimensional theory
is then still a gauged $N=1$ supergravity theory on the orbifold
$S^1/Z_2\times M_4$. It now couples to the two $N=1$ theories
on the boundary three-branes as well as to a series of $N=1$ theories
residing on the additional three-branes.

The vacuum of this five-dimensional theory is a non-trivial BPS domain
wall solution with an associated effective four-dimensional theory
which describes fluctuations around this vacuum state~\cite{Lukas:1999yy}.  It
is this four-dimensional effective theory, in its most elementary
form, which we will use as a starting point for the present paper.
First of all, for simplicity, we focus on the case of one additional
five- (three)-brane in the bulk. The generalisation of our results to
a more complicated configuration is straightforward and will be
briefly discussed in the end. For our purpose, we can truncate the
four-dimensional theory to three chiral superfields, namely the
dilaton $S$, the universal $T$ modulus and the position modulus $Z$
which specifies the position of the additional brane in the orbifold
direction. The K\"ahler potential which governs the dynamics of these
fields has been computed in Ref.~\cite{Derendinger:2001gy,AL&MB} and is given by
\begin{equation}
 K = -\ln\left( S+\bar{S}-q_5\frac{(Z+\bar{Z})^2}{T+\bar{T}}\right)
     -3\ln\left( T+\bar{T}\right)\; , \label{K}
\end{equation}
where $q_5$ is a constant.
Let us briefly explain the meaning of (the bosonic components of) these
superfields in terms of the fields in the underlying five-dimensional
theory. To do this we introduce the six real fields $(\f ,\s )$,
$(\b ,\c )$ and $(z,\z )$ through
\bea
 S &=& e^\f +q_5z^2e^\b -2i(\s -q_5 z^2\c) \nn \\
 T &=& e^\b +2i\c \label{STZ}\\
 Z &=& e^\b z-2i(\z -z\c )\nn\; .
\eea
The field $\f$ originates from the five-dimensional dilaton which is
part of the universal $D=5$ hypermultiplet. It measures the size of
internal Calabi-Yau three-fold (averaged over the orbifold) such that
the Calabi-Yau volume is given by $ve^\f$ with some fixed reference
volume $v$. The corresponding dilatonic axion $\s$ is the other
zero mode surviving from the universal hypermultiplet. 
The field $\b$ is obtained as the zero mode of the $(55)$ component
in the $D=5$ metric and as such measures the size of the orbifold.
More precisely, this size is given by $\p\r e^\b$ with a fixed
reference size $\p\r$. The associated axion $\c$ is related to
the vector field in the five-dimensional gravity multiplet.
Finally, the field $z$ originates from the world-volume of the
additional three-brane and specifies the position of this brane
in the orbifold direction. We have normalised this field to the interval
\begin{equation}
 z\in [0,1]
\end{equation}
where the two endpoints $z=0$ and $z=1$ correspond to the locations
of the boundary branes. The associated axion $\z$ also
originates from the additional three-brane and it is related to
the self-dual two-form of the underlying five-brane.
The constant $q_5$ in the above K\"ahler potential can be written
as
\begin{equation}
 q_5 = \p\e_0\b_5
\end{equation}
where $\b_5$ is the five-brane charge and $\e_0$ is defined by
\begin{equation}
 \e_0 = \left(\frac{\k}{4\p}\right)^{2/3}\frac{2\p\r}{v^{2/3}}\; .
\end{equation}
Here, $\k$ is the 11-dimensional Newton constant while $v$ and $\p\r$
have been introduced above. We note that the five-brane charge and,
hence, the constant $q_5$ needs to be positive if the compactification
is to preserve four-dimensional supersymmetry, as we have assumed.

\vspace{0.4cm}

The above K\"ahler potential is a lowest order expression in the
strong-coupling expansion parameter $\e$ defined by
\begin{equation}
 \e = \e_0e^{\b-\f}\; .\label{e}
\end{equation}
This quantity measures the size of the loop corrections to the
four-dimensional effective action, or, from a five-dimensional viewpoint,
the warping of the domain wall vacuum solution.
While the $S$, $T$ part of the K\"ahler potential does not
receive any corrections at first
order~\cite{Lukas:1998fg} in $\e$, such corrections are expected for the
$Z$-dependent part. To date, these corrections have not been computed
and we will work with the lowest order expression which is valid
as long as $\e$ is sufficiently small.

When the five-brane collides with one of the boundaries, that is, if
$z\rightarrow 0$ or $z\rightarrow 1$, one expects the appearance of new
massless states which originate from membranes stretched between the
five-brane and its $Z_2$ mirror and the five-brane and the
boundary. The theory then undergoes a small-instanton
transition~\cite{Witten:1996gx,Ganor:1996mu,Ovrut:2000qi}, where the bulk five-brane is converted 
into an
instanton (or a gauge five-brane) on the boundary. The particle
content and other properties of the $N=1$ theory on the relevant
boundary can change dramatically under this transition. At present,
the dynamics are not well enough understood to be able to follow the
cosmological evolution through this transition without additional
ad-hoc assumptions. Here, we will not attempt to improve on this
but we will merely analyse the conditions under which a brane-collision
occurs.

Perturbatively, the superpotential for the moduli fields $S$, $T$ and
$Z$ vanishes but non-trivial contributions are, of course, expected at
the non-perturbative level~\cite{Lukas:1999kt,Harvey:1999as,Lima:2001jc,Lima:2001nh}. While there is 
a long history
of analysing the phenomenology of non-perturbative superpotentials for $S$ and $T$,
particularly those induced by gaugino condensation, only recently
first efforts have been made to include the modulus 
$Z$~\cite{Khoury:2001wf,Khoury:2001iy,Kallosh:2001du}.
While this is an interesting direction for future research, here we
focus on the simplest case of a vanishing superpotential. In our
cosmological context, this amounts to finding generalisations of
rolling-radii solutions to include the position modulus $Z$. In general,
rolling-radii solutions constitute a basic and important class of
string cosmological solutions and appear to be the natural starting
point for a study of moving-brane cosmologies.

\vspace{0.4cm}

The component action of our model can be straightforwardly computed
from the K\"ahler potential~\eqref{K} together with the
reparametrisation~\eqref{STZ}. To fix our conventions, we provide the
relevant terms in the supergravity action which are given by
\begin{equation}
 S = \frac{1}{2\k_P^2}\int\sqrt{-g}\left[\frac{1}{2}R+K_{\imath{\bar{\jmath}}}
     \pt_\m\F^\imath\pt^\m\bar{\F}^{\bar{\jmath}}\right]\; .
\end{equation}
Here $K_{\imath\bar{\jmath}}=\pt^2 K/\pt\F^\imath\pt\bar\F^{\bar{\jmath}}$ is the
K\"ahler metric and $(\F^\imath)=(S,T,Z)$. In terms of
higher-dimensional quantities, the four-dimensional Newton constant
$\k_P$ can be expressed as
\begin{equation}
 \k_P^2=\frac{\k^2}{2\p\r v}\; .
\end{equation}
It can be seen that all terms in the resulting component action are
at least bilinear in the axion fields $\s$, $\c$ and $\z$. Therefore,
these fields can be consistently set to zero and we can work with
an action that is further truncated to the fields $\f$, $\b$ and $z$.
The component action then reads
\begin{equation}
\label{cmp_action}
 S = 
-\frac{1}{2\k_P^2}\int\sqrt{-g}\left[\frac{1}{2}R+\frac{1}{4}\pt_\m\f
     \pt^\m\f +\frac{3}{4}\pt_\m\b\pt^\m\b+\frac{q_5}{2}e^{\b -\f}
     \pt_\m z\pt^\m z\right]\; . \label{S4}
\end{equation}
It is for this simple action that we wish to study cosmological
solutions in the following sections. Due to the non-trivial kinetic term for
$z$, solutions with exactly constant $\f$ or $\b$ do not exist as soon
as the five-brane moves. Therefore, the evolution of all three fields
is linked and (except for setting $z=$ const) cannot be truncated
consistently any further. Recently, this has also been pointed out
in Ref.~\cite{Kallosh:2001du}. We remark that this statement remains true
in the presence of a non-perturbative superpotential.


\section{Cosmological solutions with a moving five-brane}

Let us now look for cosmological solutions to the field equations derived from
the action~\eqref{S4}. For simplicity, we assume the three-dimensional
spatial space to be flat. Our Ansatz then reads
\begin{eqnarray}
ds^2 &=& - e^{2 \nu} d \tau^2 + e^{2 \alpha} d \bx^2  \\
\phi &=& \phi(\t ) \\
\alpha &=& \alpha(\t ) \\
\beta &=& \beta(\t ) \\
z &=& z(\t )
\end{eqnarray}
Later on, we will indicate how to include three-dimensional spatial
curvature. The equation of motion for $z$ can be immediately integrated
once, leading to
\begin{eqnarray}
\label{Zeqn}
\dot z = v e^{-3 \alpha + \nu - \beta + \phi}\; ,
\end{eqnarray}
where the dot denoted the derivative with respect to $\t$.
This result may be substituted back into the rest of the system to obtain
a closed set of equations for $\alpha$, $\beta$ and $\phi$. These
equations can then be integrated in a manner closely analogous
to the formalism for anti-symmetric tensor fields in cosmology
developed in Ref.~\cite{Lukas:1997iq}. We will subsequently use this
formalism which makes the general structure of the equations and
their solutions particularly transparent. Also, several generalisations
to be discussed later are most easily incorporated in this language.

\subsection{Equations of motion and their solutions}

It turns out that the remaining equations of motion for our system
can written in the elegant form
\bea
 \frac{d}{d\t}\left( EG\bal '\right)+E^{-1}\frac{\partial 
U}{\partial\bal}
  &=&0 \label{al_eom} \\ 
 \frac{1}{2}E{\bal '}^TG\bal '+E^{-1}U &=& 0\; . \label{N_eom}
\eea
Here, the scale factors have been arranged in a vector
$ \bal = \left( \alpha, \beta,\phi \right)$ which spans the moduli
space of our cosmological model. The potential $U$ on this moduli space
is given by
\begin{equation}
U = \frac{1}{2} u^2 e^{\bq . \bal}\; , \qquad
u^2= \frac{q_5v^2}{2}\; \label{U}
\end{equation}
where the vector $\bq$ is given by $\bq=(0,-1,1)$. Further, the
``Einbein'' $E$ is defined by $E = e^{-\nu + \bd \cdot \bal}$ with
the dimension vector ${\bf d} =(3,0,0)$ and we have introduced a
metric $G={\rm diag} \left(-3,\frac{3}{4},\frac{1}{4} \right) $.
Clearly, these field equations can be obtained from the Lagrangian
\be
 {\cal L} = \frac{1}{2}E{\bal '}^TG\bal '-E^{-1}U \label{dan_lag}
\ee
which encodes the dynamics on our moduli space in a compact form. The
presence of the potential $U$ is a direct consequence of the
five-brane motion and its five-brane origin is encoded in the
characteristic vector $\bq$. In more complicated situations, for
example when spatial curvature is included, the potential $U$ consists
of more than one exponential. Even then, the
Lagrangian~\eqref{dan_lag} can sometimes be solved by employing Toda
theory. We will return to such generalisations later on.

\vspace{0.4cm}

However, for now, we are dealing with the simple potential~\eqref{U}
containing one exponential due to the moving five-brane only. This
system can always be integrated~\cite{Lukas:1997iq} and the general
solution in the comoving gauge, $\n = 0$, written in vector notation,
is given by
\begin{eqnarray}
\label{comeqn}
\bal = {\bf p}_i \ln \left| \frac{t-t_0}{T} \right| + \left( {\bf p}_f - 
{\bf p}_i \right) 
\ln \left( \left| \frac{t-t_0}{T} \right|^{- \delta} + 1 
\right)^{-\frac{1}{\delta}} + \bal_0
\end{eqnarray}
\begin{eqnarray}
\label{zeqn}
z = d \left( 1 + \left| \frac{T}{t- t_0} \right|^{- \delta} \right)^{-1} 
+ z_0\; , \label{z}
\end{eqnarray}
where $t$ is the proper time.
Let us discuss the various integration constants in this solution.
The time-scales $t_0$ and $T$ are clearly arbitrary as are the constants
$d$ and $z_0$ which parameterise the motion of the five-brane. In contrast,
the initial and final ``expansion powers'' $\bp_i$ and
$\bp_f$ each satisfy the two constraints
\begin{equation}
 \bp_n G\bp_n = 0 \;, \qquad\bp_n \cdot \bd = 1\; , \label{rr}
\end{equation}
where $n=i,f$, and, moreover, they are mapped into one another by
\begin{equation}
\bp_f = \bp_i + \delta \frac{2G^{-1}\bq}{<\bq ,\bq >} \; .\label{map}
\end{equation}
Here, the scalar product $<\cdot ,\cdot >$ is defined by
$<a,b> = {\bf a}^T G^{-1} {\bf b}$. The power $\d$ is defined by
\begin{equation}
 \d = -\bq \cdot\bp_i \; . \label{d_def}
\end{equation}
We note that the map~\eqref{map} and the conditions~\eqref{rr} are
consistent with one another in the sense that $\bp_f$ as given by
Eq.~\eqref{map} will always satisfy the constraints~\eqref{rr}
as long as the vector $\bp_i$ does. As a consequence, the six
expansion powers contained in $\bp_i$ and $\bp_f$ depend on
one independent degree of freedom. Finally, the additive constants
$\bal_0$ are subject to one further condition
\begin{equation}
 \bq \cdot \bal_0 = \ln\left(\frac{d^2 q_5 <\bq, \bq>}{8} \right)\; .
 \label{a0}
\end{equation}
Altogether, we have seven independent integration constants
as expected on general grounds.

\vspace{0.4cm}

In order for the logarithms in Eq.~\eqref{comeqn} to be well defined
the range of $t$ should be restricted to
\be
 t\in\left\{\ba{clll}
       \left[ -\infty ,t_0\right]\;
,&(-)\;{\rm 
branch} \\
       \left[ t_0,+\infty\right]\; 
,&(+)\;{\rm branch}
       \ea\right.\; .
\ee
As usual, we, therefore, find a positive and a negative time branch for
each solution. One can easily verify that the former always starts
out in a past curvature singularity while the latter ends in a future
curvature singularity.

\vspace{0.4cm}

A useful observation is that each of the above solutions is invariant
under the mapping
\begin{equation}
 \bp_i\rightarrow\bp_f\; ,\quad
 \bp_f\rightarrow\bp_i\; ,\quad
 d\rightarrow -d\; ,\quad
 z_0\rightarrow z_0+d\; .
\end{equation}
At the same time, this map implies a sign change $\d\rightarrow -\d$
as can be seen from Eq.~\eqref{map} and \eqref{d_def}. In order to
avoid having two different sets of integration constants associated to
each solution we can, therefore, fix the sign of the parameter
$\d$. It proves convenient to adopt the convention
\be
 \begin{array}{cc}\d > 0 \qquad (-) \;{\rm branch} \\
 \d < 0 \qquad (+) \;{\rm branch} \end{array} \label{conv}
\ee

\vspace{0.4cm}

Finally, for the sake of concreteness, let us present the explicit form
of our solutions by inserting the particular values for the various
general quantities which we have defined. From Eq.~\eqref{comeqn} the
scale factors can be written in component form as
\bea
 \a &=& \frac{1}{3}\ln\left|\frac{t-t_0}{T}\right|+\a_0 \\
 \b &=& p_{\b ,i}\ln\left|\frac{t-t_0}{T}\right|+
        (p_{\b ,f}-p_{\b ,i})\ln\left(\left|\frac{t-t_0}{T}\right|^{-\d}+1
        \right)^{-\frac{1}{\d}}+\b_0 \\
 \f &=& p_{\f ,i}\ln\left|\frac{t-t_0}{T}\right|+
        (p_{\f ,f}-p_{\f ,i})\ln\left(\left|\frac{t-t_0}{T}\right|^{-\d}+1
        \right)^{-\frac{1}{\d}}+\f_0
\eea
while the expression~\eqref{z} for the five-brane modulus remains
unchanged. We see that both expansion powers for the scale factor $\a$
are given by $1/3$, a fact which is expected in the Einstein frame
and can easily be deduced from the second equation~\eqref{rr}.
The initial and final expansion powers for $\b$ and $\f$ are less
trivial and are subject to the first constraint~\eqref{rr} which
explicitly reads
\begin{equation}
 3p_{\b ,n}^2+p_{\f ,n}^2=\frac{4}{3} \label{cons1}
\end{equation}
for $n=i,f$. From Eq.~\eqref{map} it follows that they are mapped into
one another by
\begin{equation}
 \left(\ba{c}p_{\b ,f}\\p_{\f ,f}\ea\right) = P
 \left(\ba{c}p_{\b ,i}\\p_{\f ,i}\ea\right)\; ,\qquad
 P = \frac{1}{2}\left(\ba{rr}1&1\\3&-1\ea\right)\; . \label{map1}
\end{equation}
We note that this map is its own inverse, that is $P^2=1$, which is
a simple consequence of time reversal symmetry. The power $\d$
is explicitly given by
\begin{equation}
 \d = p_{\b ,i}-p_{\f ,i}\; .
\end{equation}
Finally, the condition~\eqref{a0} takes the explicit form
\begin{equation}
 \f_0-\b_0 = \ln\left(\frac{2q_5d^2}{3}\right)\; .
\end{equation}

\subsection{Properties and asymptotic behaviour of the solutions}

Let us now discuss the physical properties of our solutions in some
detail. For both branches, the system behaves in a particularly
simple manner at either end of the allowed t ranges. In these limits
it is easy to see from our solution as presented in equation
\eqref{comeqn} that the 'Hubble parameters' can be written as,
\be
 \bH \equiv \dot{\bal} \simeq \frac{\bp}{t-t_0} \label{hubble}
\ee
with constant expansion coefficients ${\bf p}$ satisfying,
\be
 \bp G\bp = 0\; ,\quad \bd \cdot\bp = 1\; .\label{kk}
\ee
In view of our convention~\eqref{conv}, the expansion power $\bp$
in these relations is given by $\bp =\bp_i$ at early times (that is,
at $t\rightarrow -\infty$ in the $(-)$ branch and $t\rightarrow t_0$
in the $(+)$ branch) and by $\bp =\bp_f$ at late time (that is, at
$t\rightarrow t_0$ in the $(-)$ branch and $t\rightarrow\infty$
in the $(+)$ branch). The above relations characterise
the standard rolling radii solutions for $\a$, $\b$ and $\f$
which have been used as a basis for pre-big-bang
cosmology~\cite{Lidsey:2000mc}.
They can be obtained as exact solutions of the action~\eqref{S4}
when the five-brane is taken to be static. Their appearance as
asymptotic limits of our full solutions can be understood from
the five-brane evolution~\eqref{z}. This equation shows that,
asymptotically, the five-brane does not move and effectively
drops out of the dynamics thereby leaving us with freely rolling
radii in these limits. This demonstrates that our solutions show inflationary
behaviour at least in some region of the negative time branch.

\begin{figure}[ht]\centering
\includegraphics[height=11cm,width=9cm, angle=-90]{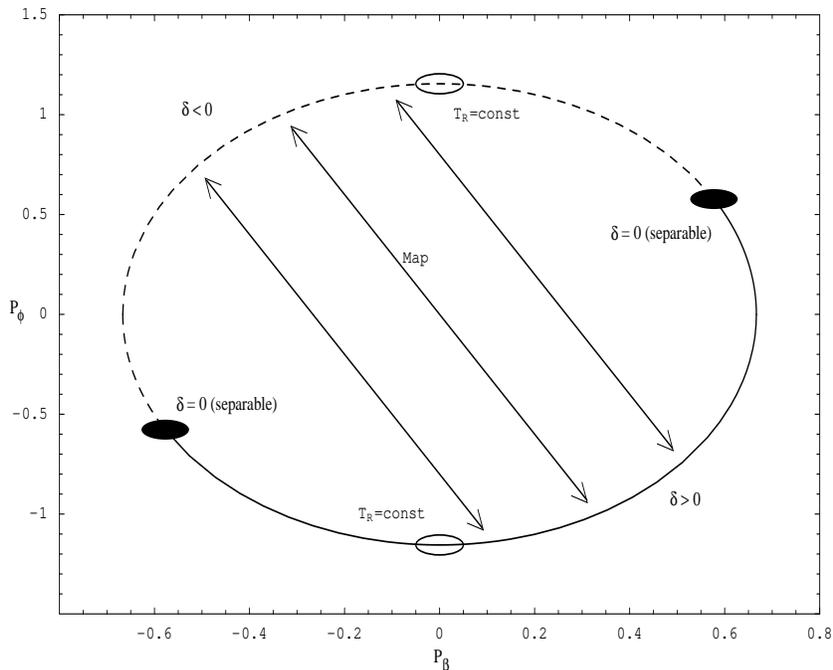}
        \caption{\emph{It is shown how the ellipse of possible
expansion powers $(p_\b ,p_\f )$ is mapped into itself under the
influence of the moving five-brane. Points related by the map may
be joined with a straight line of slope $-3$ as indicated by the arrows.
There are two fixed points under this map on the ellipse given by
$\d =p_{\beta} - p_{\alpha}=0$. These points correspond 
to solutions which are related to separating solutions in five dimensions. 
In these special cases the five-brane modulus $z$ is constant.}}
\label{fig1}
\end{figure}

Thus, we arrive at the following interpretation for our solutions. At
early time, the system starts in the rolling radii solution
characterised by the initial expansion powers $\bp_i$ while the five-brane is
practically at rest. When time approaches $|t-t_0|\sim |T|$ the
five-brane starts to move significantly which leads to an intermediate
period with a more complicated evolution of the system. Then, after a
finite comoving time, in the late asymptotic region, the five-brane
comes to a rest and the scale factors evolve according to another
rolling radii solution with final expansion powers $\bp_f$. Hence the
five-brane generates a transition from one rolling radii solution into
another one, a process which is formally described by the
map~\eqref{map1}. To illustrate the situation, in Fig.~\ref{fig1}, we
have plotted the allowed expansion powers for rolling radii solutions
in the $p_\b$--$p_\f$ plane. The arrows indicate the map between early
and late expansion powers generated by the five-brane.
\vskip 0.4cm
An obvious question to ask is whether all of the rolling radii solutions 
represented by the ellipse in Fig.~\ref{fig1} are available as
asymptotic limits at, say, early time. Perhaps surprisingly, the answer
is no. As can be seen from Eq.~\eqref{conv}, the expansion powers
at early time are subject to the constraint
\begin{equation}
 p_{\b ,i}-p_{\f ,i} =\d\;\left\{\ba{lll} >0&\quad&(-)\;{\rm branch}\\
                                        <0&\quad&(+)\;{\rm branch}
                      \ea\right. \label{conv1}
\end{equation}
Therefore only the solid (dashed) half of the ellipse is available as
the early time limit in the negative (positive) time branch. This can
also be understood from the consistency requirement that the kinetic
energy of $\b$ and $\f$ be much larger than the kinetic energy of the
five-brane modulus asymptotically if the system is to approach a rolling radii
solution. In fact, it can be easily seen that the ratio of these kinetic
energies is proportional to the strong coupling expansion parameter $\e$
defined in Eq.~\eqref{e}. For our set of solutions, the asymptotic time
evolution of this parameter is specified by
\begin{equation}
 \e\sim\left\{\ba{lll } |t-t_0|^{\d_i}&\quad&{\rm early}\\
                        |t-t_0|^{\d_f}&\quad&{\rm late}\ea\right.
 \label{e1}
\end{equation}
where we have defined
\begin{equation}
 \d_i = \d =-\bq\cdot\bp_i\; ,\qquad \d_f = -\d_i = -\bq\cdot\bp_f\; .
\end{equation}
We see that, the sign of the $\e$ expansion power is inverted as we go
from the early to the late asymptotic region. If we pick a solution
with the ``wrong'' sign, for example a $\d_i <0$ solution at early
times in the minus branch, the share of the five-brane kinetic
energy increases as $t-t_0\rightarrow -\infty$ which is inconsistent
with the rolling radii nature of the solutions in this limit.  Hence
such solutions do not exist and we are restricted to the cases which
respect~\eqref{conv1}. As a consequence, $\e$ always grows
asymptotically. This implies growing loop corrections asymptotically
or, equivalently, an increasing warping in the orbifold
direction. Hence, while there are perfectly viable rolling radii
solutions which become weakly coupled in at least one of the
asymptotic regions, the presence of a moving five-brane always leads
to strong coupling asymptotically.  A similar phenomenon has been
observed when axion dynamics is added to the simple rolling radii
set-up~\cite{Lidsey:2000mc,Copeland:1994vi}.

\vspace{0.4cm}

We note that there two exceptional solutions, indicated by the
solid dots in Fig.~\ref{fig1}, which are the fix points under the
map~\eqref{map1}. In other words, for those solutions the early and
late expansion powers are identical. It is clear from Eq.~\eqref{map}
that those solutions are characterised by $\d =0$ or, equivalently, 
by $p_\b = p_\f = \pm \frac{1}{\sqrt{3}}$. Hence, for those solutions
the orbifold and the Calabi-Yau space always evolve at the same rate
and the size of the loop corrections is time-independent. However, it
follows from Eq.~\eqref{z} that also the five-brane is static in this
case, so that these are really ordinary rolling radii solutions.
They correspond to the two separating cosmological solutions of
five-dimensional heterotic M-theory found in Ref.~\cite{Lukas:1999qs}.
In conclusion, despite these exceptional cases, it remains true
that the solutions are always driven to strong coupling
asymptotically as soon as the five-brane moves.

\vspace{0.4cm}

Our simple effective action~\eqref{S4} remains valid only
as long as loop and higher derivative corrections are sufficiently small
and this limits the validity of our solutions in the usual way.
Another, perhaps more interesting event leading to a break-down
of the effective action is the collision of the five-brane with
one of the boundaries, that is $z\rightarrow 0,1$.
In this context, an important observation from Eq.~\eqref{z} is,
that the five-brane moves a finite coordinate distance given by $|d|$, only.
Consequently, whether or not a brane collision takes place,
depends on the choices for the arbitrary integration
constants $d$ and $z_0$. This may be surprising, perhaps, as one could
have expected that a collision is unavoidable in a cosmological
setting and in the absence of any stabilising potentials.
Of course, this is related to the system always becoming strong-coupling
asymptotically which implies that the non-trivial kinetic term
for five-brane modulus leads to damping terms in the $z$ equation
of motion. We also
remark that, for each solution, the five-brane can move in either
direction, as specified by the sign of $d$. It can, therefore, collide
with either one of the boundaries. This is, of course, a direct
consequence of our simple effective action~\eqref{S4} which is
invariant under $z\rightarrow -z$ and does not single out a particular
direction.


\section{An explicit example}

We would now like to illustrate our general results by an explicit
example. For concreteness, we focus on the negative-time branch and
consider the solutions with an approximately static orbifold at early
time. In the context of our formalism, these solutions can be singled
out by setting the initial expansion power of the orbifold to zero,
that is $p_{i,\b}=0$. From Eq.~\eqref{cons1} this implies two
possibilities for the initial expansion power of the dilaton, namely
$p_{i,\f}=\pm\frac{2}{\sqrt{3}}$. Only one of these two expansion powers,
$p_{i,\f}= -\frac{2}{\sqrt{3}}$, leads to a positive value for
$\d = p_{i,\b}-p_{i,\f}= \frac{2}{\sqrt{3}}$ and can, therefore, be
realised as the early time limit of one of our solutions in the
negative time branch. Using the map~\eqref{map}, the complete set of
initial and final expansion powers for our example is then given by
\begin{equation}
 \bp_i = \left(\ba{c}\frac{1}{3}\\0\\-\frac{2}{\sqrt{3}}\ea\right)\quad
 \longrightarrow\quad
 \bp_f = \left(\ba{c}\frac{1}{3}\\-\frac{1}{\sqrt{3}}\\\frac{1}{\sqrt{3}}
         \ea\right)\; .
\end{equation}
Recall here that the first entry in those vectors specifies the
expansion power of the scale factor $\a$ (which always equals
$\frac{1}{3}$) while the second and third entry refer to $\b$ and $\f$,
respectively.
\begin{figure}[tcb]
        \centerline{\epsfbox{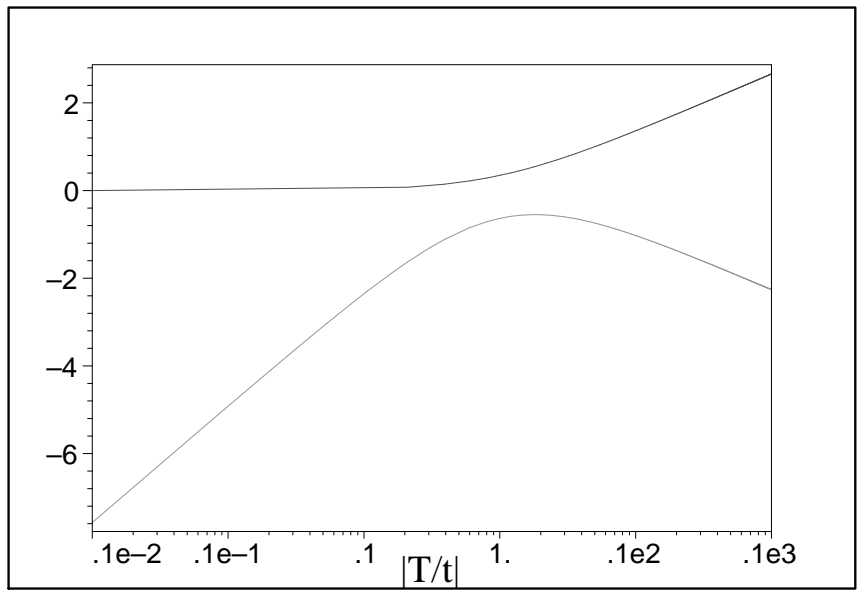}}
        \vspace{-4cm}
        \caption{\emph{Time-behaviour of $\b$ (upper curve)
        and $\f$ (lower curve) for the example specified in the text.}}
	\label{fig2}
\end{figure}
\vspace{0.4cm}

At early times, $|t-t_0|\gg |T|$,  the evolution is basically of
power-law type with powers $\bp_i$. As discussed, this happens because
at early time the five-brane is effectively frozen at $z\simeq d+z_0$
and does not contribute a substantial amount of kinetic energy. This
changes dramatically once we approach the time $|t-t_0|\sim |T|$.  In a
transition period around this time, the brane moves from its original
position by a total distance $d$ and ends up at $z\simeq z_0$. At the
same time, this changes the behaviour of the moduli $\b$ and $\f$
until, at late time $|t|\ll |T|$, they correspond to another rolling
radii solution with powers controlled by $\bp_f$. Concretely, the
orbifold size described by $\b$ turns from being approximately
constant at early time to expanding at late time, while the Calabi-Yau
size controlled by $\f$ undergoes a transition from expansion to
contraction. This behaviour has been illustrated in Fig.~\ref{fig2}.

Recall from Eq.~\eqref{e1} that the expansion powers of the strong
coupling expansion parameter $\e$ in the initial and final asymptotic
regions are given by $\d_n = p_{\b ,n} -p_{\f ,n}$, where $n=i,f$.
We have also seen in general that the transition induced by the brane
is simply inverts the sign of this expansion power, so that $\d_f=-\d_i$.
For our example, we find
\begin{equation}
 \d_i=2/\sqrt{3}\rightarrow \d_f = -2/\sqrt{3}
\end{equation}
so that $\e$ decreases initially and increases after the transition.
Consequently, the solution runs into strong coupling in both
asymptotic regions $t-t_0\rightarrow -\infty$ and $t-t_0\rightarrow 0$
which illustrates our general result.

\vspace{0.4cm}

\begin{figure}[tcb]
        \centerline{\epsfbox{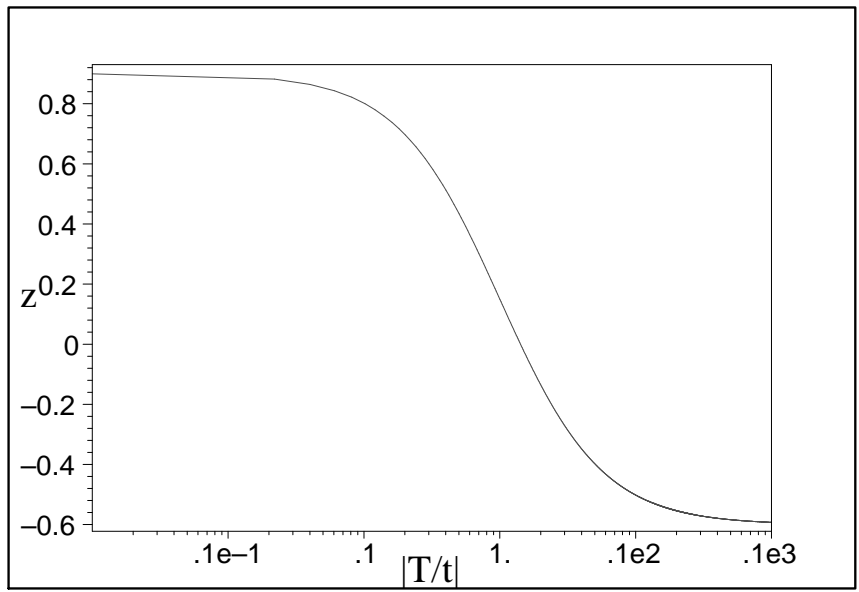}}
        \vspace{-4cm}
        \caption{\emph{Time-behaviour of the five-brane position
        modulus $z$ for the example specified in the text.
        The boundaries are located at $z=0,1$ and the five-brane
        collides with the $z=0$ boundary at $|t/T|\simeq 1$.}}
	\label{fig3}
\end{figure}

Of course, the range of validity of our solution is constrained by loop
corrections, as well as higher curvature corrections, in a way that
depends on the various integration constants involved.  If the
constants $d$ and $z_0$, characterising the five-brane motion, are
chosen such that the five-brane does not collide with the boundary there
is no further constraint. As discussed, this can always be
arranged. However, we may also choose $d$ sufficiently large so that
the five-brane does collides with one of the boundaries. Then the system
undergoes a small-instanton transition and our classical solution is
invalidated. This typically happens at the time $|t-t_0|\simeq |T|$
when the five-brane moves significantly. Depending on integration
constants this may happen well before loop and higher-derivative
corrections become relevant.
In Fig.~\ref{fig3} we have shown a particular case for our example
which leads to brane collision. The five-brane is initially located at
$d+z_0\simeq 0.9$ and moves a total distance of $d=1.5$
colliding with the boundary at $z=0$ at the time $|t-t_0|/|T|\simeq 1$.
This represents an explicit example of a negative-time branch solution
which ends in a small-instanton brane-collision.


\section{Generalisations and Outlook}

In this paper, we have presented and analysed the simplest class
of cosmological solutions with moving five-branes in the context
of four-dimensional effective heterotic M-theory. Clearly, there
are a number of interesting generalisations and refinements some of
which we would briefly like to discuss below.
\begin{itemize}
\item Including spatial curvature~: Consider the case where, in
addition to the elements included in the above discussion, we wish to
include three-dimensional spatial curvature. One of the benefits
of our general formalism is that this can be easily incorporated.
One can simply continue to work with a moduli space Lagrangian of
the form~\eqref{dan_lag}. The only change is that another
exponential term proportional to $\exp (\bq_c\cdot\bal )$ 
which accounts for the spatial curvature has to be added to the
potential $U$. The characteristic vector $\bq_c$ of this curvature term
is different from the corresponding five-brane vector and is explicitly
given by $\bq_c = (4,0,0)$. Hence, the five-brane and curvature
characteristic vectors are orthogonal with
respect to our scalar product, that is, $<\bq ,\bq_c >=0$.
Therefore, the system can be identified as an $SU(2)^2$
Toda model which can be integrated. In a different, but formally
similar context, this has been explicitly carried
out in Ref.~\cite{Lukas:1997iq}.
\item Add a perfect fluid~: One may want to consider the cosmology of
our simple model in the additional presence of a perfect fluid.
As before, this can be done by simply adding an exponential term
to the potential $U$. The characteristic vector for a perfect
fluid with equation of state $p=w\r$, where $w$ is a constant, is given by
$\bq_{\rm fluid}=(3(1-w),0,0)$ which is again orthogonal to the five-brane
vector. Hence this situation is again described by an
$SU(2)^2$ Toda model.
\item More than one five-brane~: We may also easily generalise our
results to include more than one five brane. This simply involves
including more potential terms with $\bq = (0,-1,1)$. This, of course,
simply amounts to replacing the expression for the normalisation $u^2$
of the potential $U$ in Eq.~\eqref{U} by a generalisation which involves
summing over all five-branes. Most of the qualitative features of
our solutions will then be unchanged. In particular,
all five-branes 'move at the same time' collectively generating the
transition between the two asymptotic regions. However, the
initial position and the total distance of the motion can
be different for the various five-branes. For example, some of the
five-branes may collide with a boundary while the others do not. Of course the five-branes may also 
collide with one another.

\item Include axions~: Another obvious generalisation to attempt
would be to try and include  non-trivial axion dynamics in our
solutions. In more simple string cosmology settings without five-branes
this has been possible due to an $SL(2)$ symmetry of the string 
effective action \cite{Lidsey:2000mc,Copeland:1994vi}. This symmetry allowed solutions with
non-trivial axion dynamics to be generated from systems where the
axion fields had been set to zero. However, the presence of the
five-brane modulus $Z$ complicates the situation significantly.
The only parts of $SL(2)$ which are obviously respected by the
K\"ahler potential~\eqref{K} are the axionic shift symmetries
which do not lead new solutions. It is not clear, at present,
whether the K\"ahler potential~\eqref{K} has any further symmetries
which could be used to generate new cosmological solutions from
the ones obtained in this paper.
\item Small instanton transition~: Finally, an understanding of the
dynamics of the small instanton transition whereby a five brane is
emitted from or absorbed onto one of the fix point branes would be
extremely useful. It would allow us to describe the system as it
evolves through the transition which is vital in order to construct
``complete'' cosmological models.
\item Fluctuations~: It is known~\cite{Copeland:1997ug,Lidsey:2000mc}
      that there exist special rolling radii solutions which, in the context
      of pre-big-bang inflation, lead to perturbations consistent
      with a Harrison-Zeldovich spectrum. It is, therefore, expected
      that such cases with
      an acceptable fluctuation spectrum also exist in our generalised
      framework with moving five-branes. We will be analysing this
      in more detail.
\end{itemize}

Another possible direction for future work would be a search for five
dimensional moving brane solutions of various kinds, including those
that correspond to the four dimensional ones found here. It is,
of course, straightforward to ``oxidise'' all our four-dimensional
solutions to approximate solutions of five-dimensional heterotic
M-theory following the procedure described in Ref.~\cite{Lukas:2000yn}.
However, finding the exact solutions in five dimensions in the
presence of additional five-branes is a much more difficult task.
We have seen that the five-brane is static for the two exceptional
solutions with $\d =0$. In this sense, they are trivial extensions
of $D=4$ rolling radii solutions. Finding their $D=5$ counterpart,
however, is less trivial, since the presence of the additional
five-brane changes the warping of the $D=5$ domain wall, an effect
which is not directly visible in $D=4$. However, one can verify that
the two separating $D=5$ cosmological solutions found in
Ref.~\cite{Lukas:1999qs} can indeed be generalised to include five-branes.
Hence, the $D=5$ counterparts of our two exceptional solutions can
be obtained as exact solutions. More generally, we have shown
that there exist only three $D=5$ solutions, based on the bulk
BPS domain wall solution of Ref.~\cite{Lukas:1999yy}, which include a single
five-brane. These are the BPS domain wall solution with a single five-brane
itself and the two aforementioned cosmological solutions. 
Hence, lifting the other solutions we have presented to exact $D=5$
solutions is not straightforward and requires the study of more
complicated metrics in the bulk such as those found in
Ref's~\cite{Chamblin:1999ya,Feinstein:2001xs}.


\vspace{1cm}

\noindent
{\Large\bf Acknowledgements}\\
A.~L.~is supported by a PPARC Advanced Fellowship. J.~G.~ is supported
by PPARC.



\end{document}